\documentclass[journal=jacsat,manuscript=article]{achemso}

\usepackage[version=3]{mhchem} % Formula subscripts using \ce{}

\usepackage{color}

\author{Liangzhi Kou}
\affiliation{Bremen Center for Computational Materials Science, University of Bremen, Am Falturm 1, 28359 Bremen, Germany}
\email{kouliangzhi@gmail.com}
\author{Shu-Chun Wu}
\affiliation{Max Planck Institute for Chemical Physics of Solids, Noethnitzer Str. 40, 01187 Dresden, Germany}
\author{Claudia Felser}
\affiliation{Max Planck Institute for Chemical Physics of Solids, Noethnitzer Str. 40, 01187 Dresden, Germany}
\author{Thomas Frauenheim}
\affiliation{Bremen Center for Computational Materials Science, University of Bremen, Am Falturm 1, 28359 Bremen, Germany}
\author{Changfeng Chen}
\affiliation{Department of Physics and Astronomy and High Pressure Science and Engineering Center, University of Nevada, Las Vegas, Nevada
89154, United States}
\author{Binghai Yan}
\affiliation{Max Planck Institute for Chemical Physics of Solids, Noethnitzer Str. 40, 01187 Dresden, Germany}
\alsoaffiliation{Max Planck Institute for the Physics of Complex Systems, Noethnitzer Str. 38, 01187 Dresden, Germany}
\email{yan@cpfs.mpg.de}

\title[An \textsf{achemso} demo]
{New Family of Robust 2D Topological Insulators in van der Waals Heterostructures}

\abbreviations{IR,NMR,UV}
\keywords{American Chemical Society, \LaTeX}

\begin{document}

\begin{tocentry}
 \centering
\includegraphics [width=5cm]{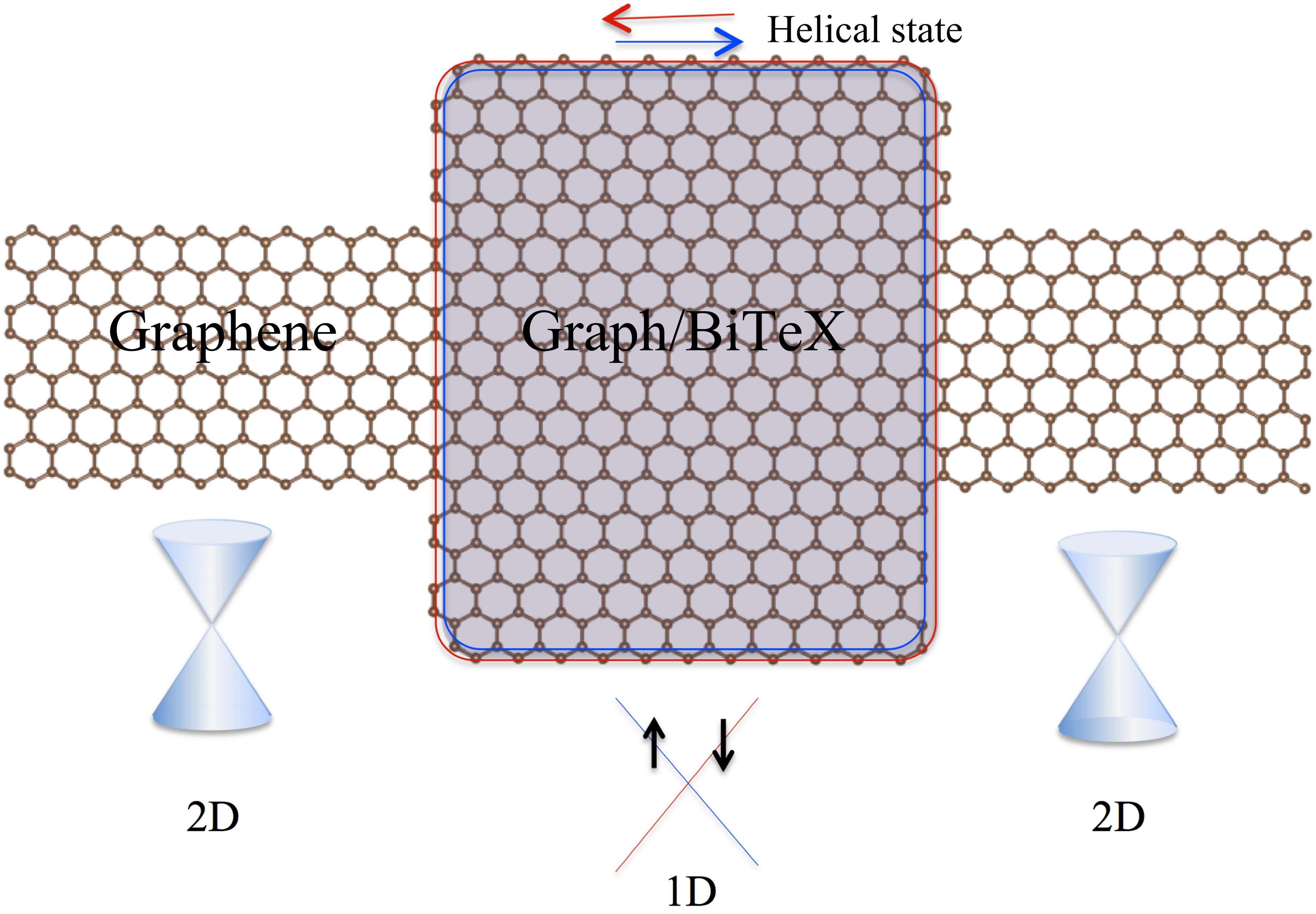}
\end{tocentry}

\begin{abstract}
We predict a new family of robust two-dimensional (2D) topological insulators in van der Waals heterostructures comprising graphene and chalcogenides BiTeX (X=Cl, Br and I).  The layered structures of both constituent materials produce a naturally smooth interface that is conducive to proximity induced new topological states.  First-principles calculations reveal intrinsic topologically nontrivial bulk energy gaps as large as $70-80$ meV, which can be further enhanced up to 120 meV by compression.  The strong spin-orbit coupling in BiTeX has a significant influence on the graphene Dirac states, resulting in the topologically nontrivial band structure, which is confirmed by calculated nontrivial $Z_2$ index and an explicit demonstration of metallic edge states.  Such heterostructures offer an unique Dirac transport system that combines the 2D Dirac states from graphene and 1D Dirac edge states from the topological insulator, and it offers new ideas for innovative device designs.
\end{abstract}

\section{Introduction}

Topological insulators (TIs) are a new quantum state of matter that exhibits robust metallic boundary states inside the bulk energy gap \cite{1,2,3,4}.  TIs have been discovered in many materials such as chalcogenides (e.g., HgTe \cite{5,6} and Bi$_2$Se$_3$ family \cite{7,8}) and oxides (e.g., BaBiO$_3$ \cite{9}), and they have inspired great research interest in physics, chemistry and materials science~\cite{10,11} to explore their promising potential for applications in spintronics, thermoelectrics and quantum computation \cite{3,4}. For example, the topological edge states of a 2D TI carry dissipationless current because of the robustness against backscattering, and this may lead to low-power consumption electronic devices. TIs can also be realized at the interface between two different materials. For instance, InAs/GaSb quantum wells (QWs) \cite{12,13} were found to be 2D TIs due to the inverted order between the InAs conduction and GaSb valence bands. Recently more 2D TIs were also predicted in serval systems similar to InAs/GaSb QWs \cite{14,15,16}. The existence of TI states in these systems requires an atomically smooth interface region, which presents a challenge to the experimental fabrication.

A heterostructure between two naturally layered materials tend to form a smooth interface connected by weak interlayer van der Waals (vdW) type interactions.  Graphene-based vdW heterostructures have been fabricated recently, and they exhibit interesting properties and new phenomena \cite{17}.  Spin-orbit coupling (SOC) and superconductivity can be induced from one layer to the other via proximity effect at the vdW interface.  For example, the hybrid structure of a single layer of Bi$_2$Se$_3$ and a graphene layer possesses an inverted band driven by the strong SOC of the Bi$_2$Se$_3$ layer, producing a 2D TI \cite{18};  proximity induced superconductivity has been observed in the MoS$_2$ layer by NbS$_2$ \cite{19,20}, resulting in 1D topological superconductivity that is an ideal platform for realizing Majorana fermions for quantum computation applications \cite{21,22}. In addition, graphene-based TIs with sizable energy gaps are a long-sought goal since the advent of TIs \cite{k23}.  For this purpose, adatom deposition on graphene was proposed \cite{24,25,26}. However, this method is usually vulnerable to non-homogeneous adatom coverage or induces unwanted electron doping \cite{24}. In contrast, a vdW interface can avoid these problems.  As a result, vdW heterostructures may hold great promise for designing new graphene based topological materials.

In this Letter, we explore a vdW heterostructure between the graphene and chalcogenide BiTeX (X=Cl, Br and I) monolayer using first-principles calculations, and we predict that they form a 2D TI with an energy gap as large as $70-80$ meV. The chalcogenide layer does not need to be a TI, and its role is to induce a strong SOC to modify the graphene Dirac bands.  To maximize the proximity effect, the graphene layer is sandwiched between two BiTeX layers in a quantum well (QW) configuration, in which the graphene Dirac bands hybridize strongly with the Bi- and Te-$p$ states, resulting in a large energy gap.  The interplay between these two materials in the QW produces a nontrivial band structure with a topological invariant $Z_2 = 1$, and the nontrivial topological state is further confirmed by an explicit demonstration of the presence of gapless edge states from our calculations. External pressure within the experimentally accessible range can further increase the energy gap up to more than 100 meV. The atomic configurations of QWs (e.g.,  the layer stacking sequence and thickness) also influence the resultant electronic properties.

\section{Methods}

First-principles calculations based on the density functional theory (DFT) were carried out using the Vienna Ab Initio Simulation Package (VASP) \cite{27}. The exchange correlation interaction was treated within the local-density approximation (LDA). The QW is periodic in the $xy$ plane and separated by at least 10 {\AA} along the $z$ direction to avoid the interactions between adjacent QWs. All the atoms in the unit cell are fully relaxed until the force on each atom is less than 0.01 eV/{\AA}. The Brillouin-zone integration was sampled by a 14 $\times$ 14 $\times$ 1 $k$-grid mesh. An energy cutoff of 400 eV was chosen for the plane wave basis. The SOC was included. To describe the vdW interaction, a semiempirical correction by Grimme method \cite{28} was adopted. The DFT Bloch wave functions were further projected to maximally localized Wannier functions \cite{29} for the edge state calculations.

\section{Results and discussion}

\begin{figure}
\centering
\includegraphics [width=8cm]{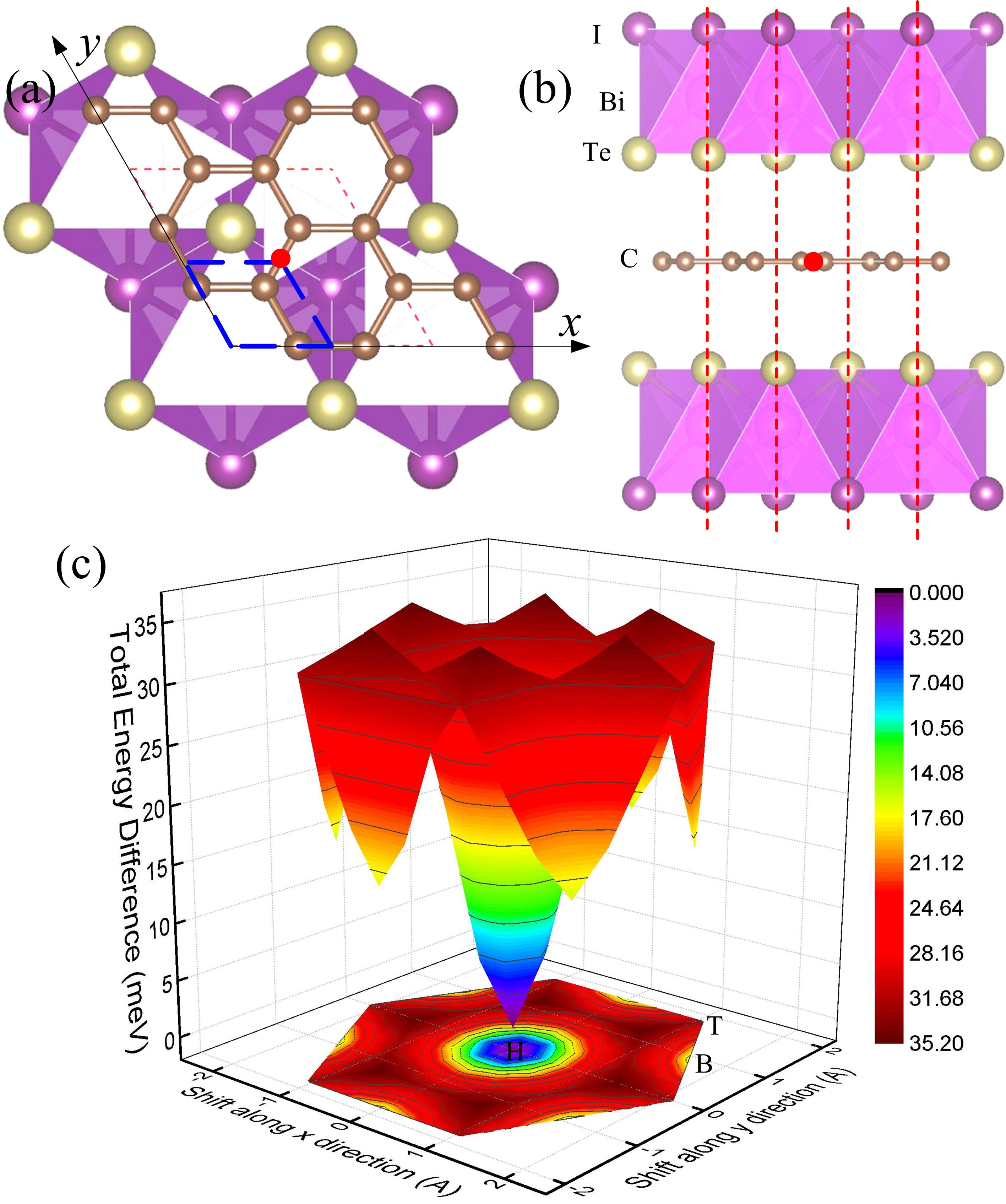}\\
\caption{Top (a) and side (b) view of the structural model of the most stable configuration for the BiTeI/Graphene QW, where the unit cell is indicated by the dashed red lines.  Two BiTeI layers sandwich a graphene layer where the Te atoms of the BiTeI layers locate at the centers of the graphene hexagons. Inversion symmetry is preserved inside the QW with the inversion center at the middle of the C-C bond, as indicated by the red dot. (c) The total energy variation dependence on the stacking position of the BiTeI layer.  The energy of the QW with Te at the carbon hexagon center is set to zero as a reference. The layer shifting coordinates in the $xy$ plane are with respect to where the Te atoms match the hexagon center, which is the most stable position.  One third of the total shifting region is indicated by the blue dashed lines in (a). H, B and T represent the Te atoms located at the hexagonal hollow, C-C bridge and on carbon top positions, respectively. The interlayer distance between the graphene and BiTeI layers is predicted to be about 3.3 {\AA}, which is in the reasonable range of vdW interactions.}
\label{Fig. 1}
\end{figure}

In the QW structure a graphene layer is sandwiched between two BiTeI slabs, as shown in Figs. 1(a) and 1(b). Here we choose the BiTeI monolayer as the cladding layers of the QW structure.  The experimental in-plan lattice constant of BiTeI is 4.34 {\AA} while the lattice constant of graphene is 2.46 {\AA}. We thus choose 2.46$\times\sqrt{3}$=4.26 {\AA} as an effective lattice constant, where an  in-plan supercell of $\sqrt{3}\times\sqrt{3}$ for graphene and the unit cell of BiTeI are used.  In this case, the lattice mismatch is only 1.8\%, which may lead to a slight band gap modulation of the semiconducting BiTeI, but it is not expected to affect our main conclusions.  While graphene has a single atomic layer structure, the BiTeI layer lacks inversion symmetry with its three-atomic-layer stacking in a sequence of I-Bi-Te along the $z$ axis. Due to the strong covalency and ionicity of the Bi-Te and Bi-I bonds, respectively, the triple layer possesses an intrinsic polar axis along the $z$ direction. Despite the strong chemical bonding within each triple layer, the adjacent triple layers are only weakly coupled via the vdWs interaction \cite{30,31}.  Since the two surfaces of the BiTeI layer are terminated with different atoms and the stacking patterns between the graphene and BiTeI layers can vary, it is crucial to first determine the most stable configuration of teh QW structure.  To achieve this goal, the problem is divided into two steps.  We first check the stacking preference of a bilayer structure containing one BiTeI layer and one graphene layer within two models (see Figure S1a-1d; I-Bi-Te-C and Te-Bi-I-C, which is written according to the atomic sequence along the $z$ axis), and then shift the BiTeI layer relative the graphene layer to search for the configuration with the lowest total energy.  The calculations indicate that in either case, the configuration is the most stable when the adjacent atom of the BiTeI layer (i.e., Te or I) locates at the hexagon center of the graphene layer, but all the structures in I-Bi-Te-C configurations have lower total energies (by about 20 meV per unit cell) than those in the structural model with the Te-Bi-I-C sequence (see Figure S1e).  These results show that the configuration with the Te atom of the BiTeI layer located next to the graphene layer and at the hexagon center is the most stable. At the second step, an additional cladding layer is added on the other side of the graphene layer, and here the most stable configuration with the Te atom adjacent to the graphene layer is used.  We then preformed a high throughput structure search to check all possible stacking patterns at the interface between the graphene and BiTeI layers.  The additional BiTeI layer is initially placed such that the Te atoms locate at the hollow center of the carbon hexagonal rings, and it is then moved within the half unit-cell region (see the dashed blue region, which covers all possible stacking pattern) while keeping the bilayer stacking of I-Bi-Te-C obtained in the first step unchanged.  The total energy variations for all the configurations as a function of shifted distances are shown in Fig. 1(c).  One can clearly see that the structure with the Te atoms located at the center of the carbon hexagonal rings is the most stable, while those with the Te atoms on top of the carbon atoms have the highest total energy (about 35 meV higher per unit cell). To compare QWs with different stacking patterns, we have examined three structural models with the atomic sequence along the $z$ direction as I-Bi-Te-C-Te-Bi-I, I-Bi-Te-C-I-Bi-Te and Te-Bi-I-C-I-Bi-Te following the dipole directions.  In all these cases, the atoms of the BiTeI layer adjacent to the graphene layer are all located at the hexagon centers due to the lowest energy.  The total energy of the I-Bi-Te-C-Te-Bi-I configuration is found to be 60 meV lower than the Te-Bi-I-C-I-Bi-Te configuration and 26 meV lower than the Te-Bi-I-C-Te-Bi-I configuration.  In the following, we focus our investigations on this most stable QW structure, which is shown in Fig. 1(a) and 1(b).  For brevity, the BiTeI/Graphene/BiTeI QW will be labeled as BiTeI/Graph in the following discussions.  It should be noted that this QW possesses structural inversion symmetry between the two cladding slabs, which avoids the Rashba splitting of the graphene bands ~\cite{32}.

To set a benchmark for comparison, we first establish the electronic band structure for the graphene and BiTeI double layers without introducing a coupling between them. The calculated results are shown in Fig. 2(a).  The freestanding graphene layer with a $\sqrt{3}\times\sqrt{3}$ supercell has two Dirac valleys at the K and K$^{\prime}$ points that are folded onto the $\Gamma$ point in the Brilloun zone with a zero gap.  The BiTeI double layer exhibits an indirect band gap of 0.72 eV with significant Rashba SOC splittings in both the conduction and valance bands \cite{30,31,33}.
These results are consistent with previous reports \cite{34}.  When the graphene layer is sandwiched between two BiTeI layers, the band structure of the QW is basically a superposition of the two constituent components when the SOC is not included. The low energy region of the graphene band structure near the Fermi level retains a linear band dispersion. The Dirac fermions, however, is no longer massless since a small band gap of 8.7 meV has opened up (Fig. 2b) by the intervalley scattering caused by the substrates \cite{35}.  Only in energy regions far away from the Fermi level, there are more significant band hybridizations. For instance, in the valance band around -0.6 eV, a band state from the hybridization of the C \emph{p}$_z$ with the Te-5\emph{p} and I-5\emph{p} of BiTeI appears.

\begin{figure}
\centering
\includegraphics [width=16cm]{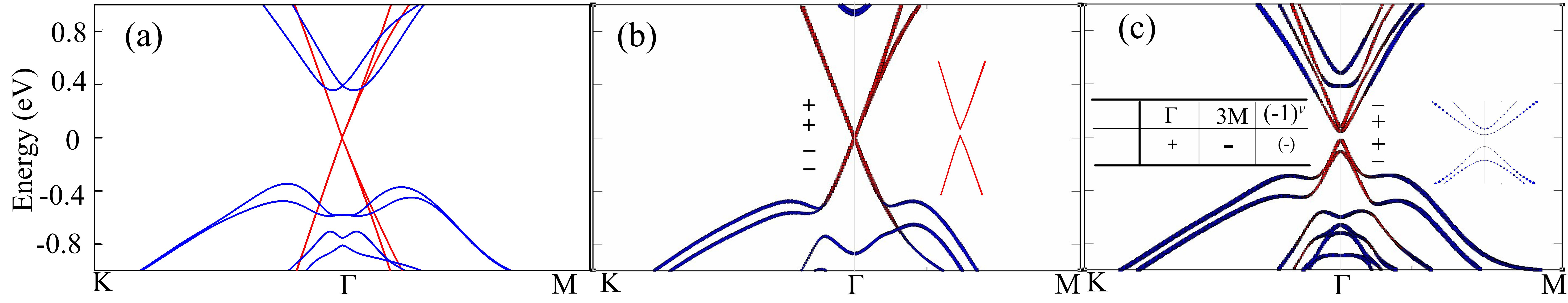}\\
\caption{ (a) Calculated band structures of the BiTeI/Graph QW (a) with the SOC but without the coupling between the graphene and BiTeI layers, (b) without the SOC but with the interlayer coupling, and (c) with both the SOC and interlayer coupling.  The red (blue) dots represent the contributions from graphene (BiTeI). The valence and conduction bands near the Fermi energy are zoomed in as insets in (b) and (c). The parities at $\Gamma$ and M are listed. Parity values for top two valence bands and bottom two conduction bands are presented in both (b) and (c) insets, where a parity change for the bands without and with the SOC is observed.}
\label{Fig. 2}
\end{figure}

The strong spin-orbit interaction and the associated Rashba effect in BiTeI lead to significant band splitting, and they are also expected to strongly influence the electronic properties of graphene and the interaction between the graphene and BiTeI layers through proximity effect \cite{36}. Our calculations verify these expectations.  With the SOC switched on, the Dirac bands of the graphene layer remain inside the gap of the BiTeI slabs, but there is a band splitting near the Fermi level, which is attributed to the fact that the induced SOC by adjacent BiTeI slabs has an effective repulsive interaction for the folded Dirac bands at the $\Gamma$ point.  As a result, the conduction bands of the BiTeI layers are significantly shifted downwards (the blue dots in Fig. 2c), and these bands hybridize with the Dirac states (low-energy conduction bands) of graphene near the Fermi level (the red dots in Fig. 2c).  Meanwhile the valance bands of BiTeI are shifted upwards and hybridize with the valance bands of graphene near the Fermi level. A band composition analysis indicates that most of the band states near the Fermi level at the $\Gamma$ point are contributed by the graphene layer, and there is only a slight contribution from the BiTeI layers (see the inset in Fig. 2c).  The influence of the BiTeI layers remarkably increases the intrinsic SOC of graphene, which brings the two Dirac cones into close proximity and then reopens the gap after inducing a band inversion.  It produces a topological phase transition in the graphene layer, rendering it a true 2D TI. The opened nontrivial band gap reaches up to 70 meV, which is well above the energy scale of room temperature \cite{7} and represents a four-orders-of-magnitude enhancement over the intrinsic gap of pristine graphene \cite{37}.  While the BiTeI layers play a key role in generating the proximity enhanced TI band gap in the graphene layer, most of the electronic bands near the Fermi level are from the graphene layer itself, making the QW structure a desirable system for study of the quantum spin Hall (QSH) effect in graphene.  The fundamental TI nature is captured by the parity of the electronic states measured by the Z$_2$ index due to its structural inversion symmetry, with four time-reversal invariant points in the 2D Brillouin zone (BZ), namely one $\Gamma$ (0, 0) point and three M (0.5, 0) points, where the coordinates are in units of the in-plane reciprocal lattice vectors. The $Z_2$ topological invariant is calculated by the product of parity eigenvalues for all the valence bands following the Fu-Kane parity criteria \cite{38}. The total parity product at the $\Gamma$ and $M$ points is ``$-$'', thus confirming that the QW structure is a 2D TI with a $Z_2$ index $\nu=1$.

Since the enhanced SOC in graphene is induced by the proximity effect, i.e., the hybridization between the graphene and BiTeI layers, it is expected that a reduction in the interlayer distance would lead to further enhanced SOC.  In experiment, this can be achieved by applying a compressive pressure perpendicular to the QW surface. To verify this scenario, we have simulated the effect of compression by reducing the distance between the outmost iodine atom and the graphene layer, and then relaxing the structure while fixing the \emph{z} coordinates of the two outmost iodine atoms (see the inset in Fig. 3).  The definition of the applied strain of the BiTeI/Graph QW is the relative distance change between the two outmost iodine atoms over the thickness of the QW.  The compression pushes the layers in the QW structure closer, which, in turn, enhances the hybridization between the BiTeI and graphene layers.  Results in Fig. 3 show that the nontrivial band gap $\Delta$ increases almost linearly from 70 meV to 100 meV (about 1,200K) at an interlayer distance compression of 0.3 {\AA} with a strain of 4.3\%, corresponding to a nominal pressure of 4 GPa, which is estimated as the energy derivation per unit area over the reduced distance.  The parity check indicates that the vertical strain increases the gap opening without changing the band topology, and the strained QWs are still 2D TIs.  This controllability of the nontrivial gap in the QW provides a freedom for electronic modulation of the 2D TI.  The band structure of the QW under pressure indicates that the hybridization of graphene and BiTeI states is enhanced, which further increases the SOC of graphene, yielding the increase of the nontrivial gap.  Further increase of pressure will push the Dirac cone into the valance band of the BiTeI layer, thus changing the band topology. The nontrivial band structure in the present vdWs heterostructure QW with or without the external compression is considerably larger than those in previously realized QSH systems such as HgTe  \cite{5,6} and InAs/GaSb ~\cite{12,13} QWs.  This result points to a feasible and promising approach for new 2D TI design.

The thickness of the BiTeI cladding layers in the QW structure may affect the electronic properties and band topology due to the presence of the dipole at each side. To address this issue, we have studied the effect of muliti-layers of BiTeI on  the electronic properties of the QW (see Supporting Materials).  When one additional BiTeI layer is added on each side, the dipole effect causes a significant upward shift of the BiTeI valance bands, which is similar to the situation at the [0001] ZnO surface \cite{39}.  For such double-layer BiTeI/Graph QW, the parity product of all the valance bands at the four time-reversal invariant points in the 2D BZ is unchanged at  Z$_2$=1, although an "$M$"-shaped state forms near the $\Gamma$ point.  The gap at the $\Gamma$ point is reduced to 12 meV as a result [Fig. S2(a)].  When a third BiTeI cladding layer at each side is added, the upward shifted BiTeI valance bands hybridize with the graphene conduction bands, and the above physical picture regarding the SOC enhancement is no longer valid.

\begin{figure}
\centering
\includegraphics [width=10cm]{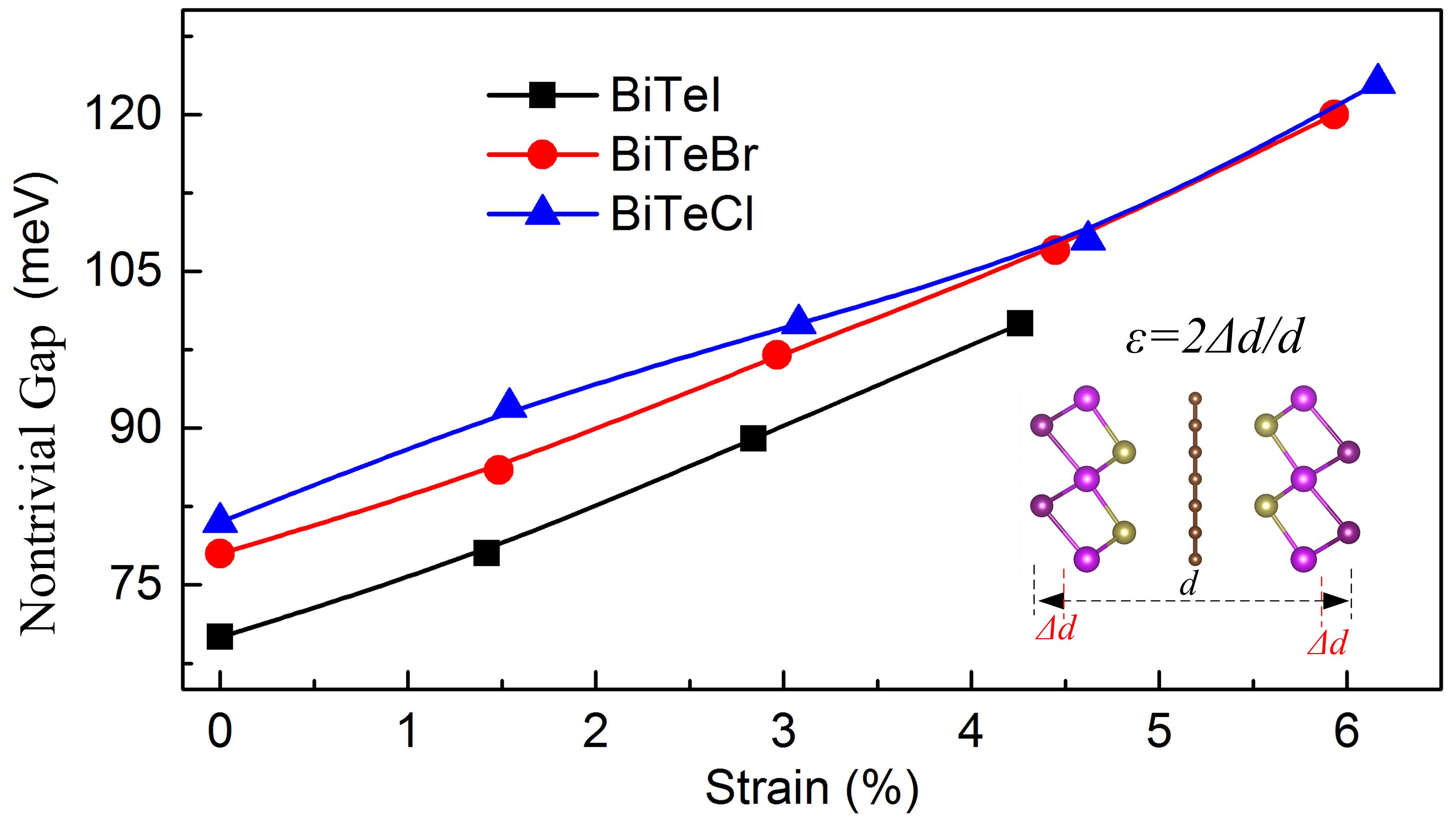}\\
\caption{Band gaps of the BiTeX/Graph QWs as a function of the applied vertical strain.  The largest strain studied here corresponds to a pressure of about 4 Gpa, which is within the experimentally accessible range.}
\label{Fig. 3}
\end{figure}

We have performed calculations to examine the topological edge states on the boundary of the QWs since these are the characteristic features and direct evidence of the 2D TI and QSH states.  Using tight-binding parameters based on the Wannier orbitals extracted from the DFT calculations, we constructed a model ribbon structure of the BiTeI/Graph QW that is 50-BiTeI-unit wide.  The graphene layer exhibits an armchair-type structure on the boundary of the ribbon. The band structure was obtained by a direct diagonalization of the tight-binding Hamiltonian of the ribbon along the 1D Brillouin zone. As shown in Fig. 4, a pair of 1D Dirac states exist inside the energy gap of the QW.  Because of the inversion symmetry, the edge states from the two ribbon edges are exactly degenerate in energy. At a given edge, two counter-propagating edge states exhibit opposite spin polarizations, a typical feature of a QSH system.

\begin{figure}
\centering
\includegraphics [width=8cm]{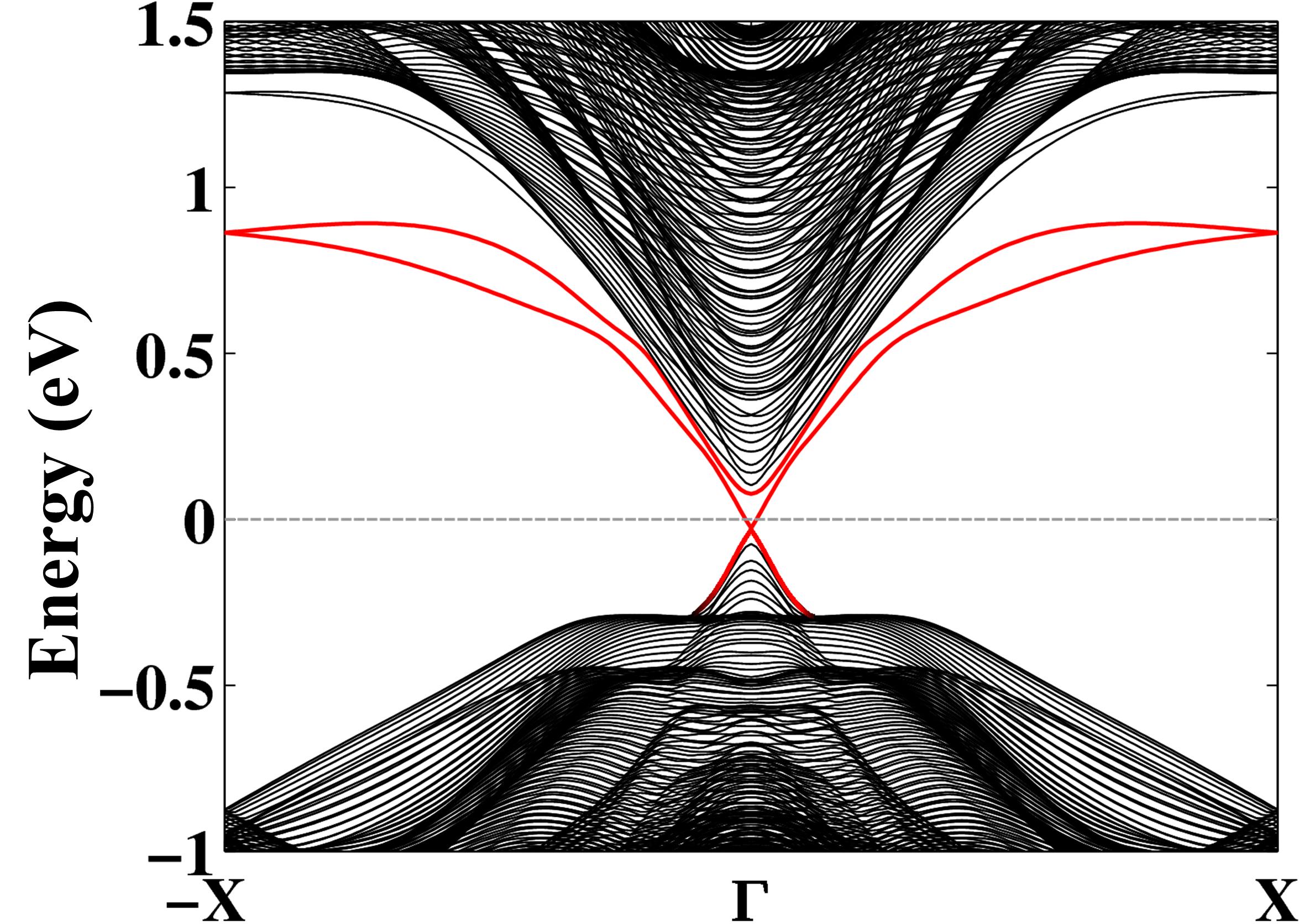}\\
\caption{The helical edge states of the BiTeI/Graph QW. The Dirac helical states are denoted by the red solid lines, which exist at the edges of the ribbon structure.}
\label{Fig. 4}
\end{figure}

From the Z$_2$ index and edge state calculations, we have confirmed that the BiTeI/Graph QW is a new 2D TI.  On a more general ground, we can also understand the underlying physics in an intuitive way.  BiTeI is a semiconductor with a large trivial band gap without a band inversion between the conduction band Bi-6p and the valance band Te-5p states, but it possesses a strong crystal-field splitting and the Rashba effect \cite{31}. The introduction of the Dirac state of graphene near the Fermi level, which is located inside the BiTeI gap, provides a platform for engineering a band inversion and topological phase transition when the intrinsic SOC in the graphene layer is greatly enhanced by the proximity effect from BiTeI. From this analysis, we can expect the same physical phenomenon when replacing the cladding layers of BiTeI with other semiconducting materials with strong SOC or Rashba effect.  To demonstrate this point, we also investigated the QWs of BiTeCl/Graphene/BiTeCl and BiTeBr/Graphene/BiTeBr, where both BiTeCl and BiTeBr are not TIs but exhibit strong Rashba effect.  A high through-output search for stable configurations shows that the structure with the Te atoms located at the carbon hexagonal centers is the most stable configuration for both cases, similar to the result for the BiTeI/Graph QW shown in Figure 1(a) and 1(b). Electronic band structure calculations and Z$_2$ index check demonstrate that the BiTeCl/Graph and BiTeBr/Graph QWs are also 2D TIs with nontrivial band gaps of  81 meV and 78 meV, respectively (see Supporting Materials).  Under a vertical compression, the nontrivial gap can reach up to 120 meV at a distance reduction of 0.4 {\AA}, corresponding to a nominal pressure of about 4 GPa (see Fig. 3).  This giant nontrivial gap of 120 meV makes the present QW structures suitable for practical nanodevice applications.  When additional cladding layers are added to each side, similar changes in the electronic band structure are obtained as seen in the BiTeI/Graph QW (see Fig. S3).

The DFT calculations (LDA or GGA) are known to usually underestimate the band gap.  To address this issue, we have performed additional calculations for BiTeI/Graph QW using hybrid functionals (HSE06), and obtained almost identical (to the LDA results) electronic band structure, especially the band gap value and band topology.  This is attributed to the fact that the nontrivial gap opening in the QW is caused by the greatly enhanced SOC from the proximity effect that is driven by the relativistic effects, which is insensitive to the choice of the exchange functionals.  It indicates that our LDA calculations can capture the main physics underlying the topological nature of the BiTeX/Graph QWs.

\begin{figure}
\centering
\includegraphics [width=12cm]{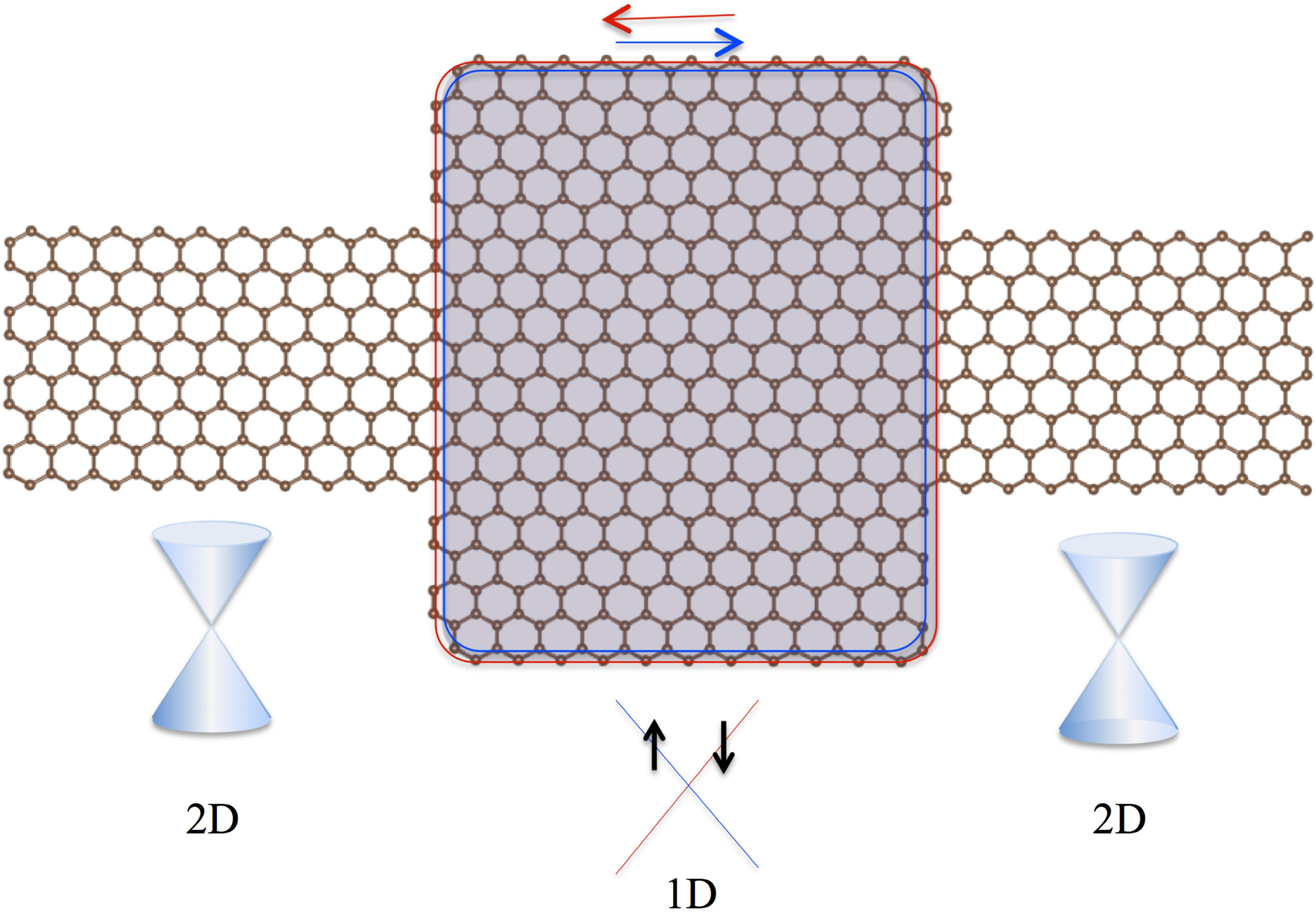}\\
\caption{Proposed device design for novel Dirac transport, where the helical spin states are indicated along the edge with the red lines for spin-up electrons with counterclockwise transport while blue line for spin-down electrons with clockwise transport.  The shaded region is coved with the BiTeX cladding layers.}
\label{Fig. 5}
\end{figure}

The unique topological features of the electronic band structure of the BiTeX/Graph QWs unveiled in the present work suggest novel device designs.  We show in Fig. 5 an illustration of a generic composite system containing a graphene component and a QW component that can be constructed by adding the BiTeX cladding layers in any designed pattern.  The central idea here is to form regions where the helical edge states can be introduce in a controlled way in a graphene template, and such states move along the patterned edges of the QW, which can be used as dissipationless conducting wires for electronic circuits.  The helical states combined with the outstanding physical properties in graphene may lead to devices with significantly lower power consumption and heat generation, which is crucial for the development of new generations of integrated circuits.  More important and interesting, such a heterostructure device can simultaneously host 2D Dirac states from graphene and 1D Dirac edge states from the 2D TI, thus offering an unique  Dirac transport system, which can be exploited for innovative applications.

Finally, we comment on possible fabrication routes for the proposed TI QW, which are formed between graphene, which has a 2D planar structure, and BiTeX, which has a trilayer structure, and they couple through weak vdW interactions.  Ultrathin layered BiTeX may be obtained by micromechanical cleavage and exfoliation methods \cite{40}.  A feasible approach to large-scale production of the vdWs heterostructures may involve growing individual monolayer graphene on catalytic substrates, then isolating and stacking with exfoliated BiTeX layers.  This route has already been proved to be scalable \cite{41}. An alternative way is to grow graphene on the BiTeX surface, or grow BiTeX using graphene as a substrate, and the feasibility of this approach is supported by the successful fabrication of graphene/Bi$_2$Se$_3$ heterostructures using vapor-phase deposition and molecular beam epitaxy techniques \cite{42,43}.

In summary, we show by first-principle calculations that BiTeX/Graph (X=I, Cl, Br) QWs constitute a new family of robust 2D TIs. The strong SOC in the BiTeX layers has a significant influence via proximity effect on the graphene Dirac states near the Fermi level, producing a topologically nontrivial state.  The nontrivial topological characteristic is confirmed by our calculated Z$_2$ index and an explicit demonstration of the existence of the helical edge states.  A large nontrivial bulk gap of 70-100 meV has been obtained, which represents a four-orders-of-magnitude enhancement compared with the intrinsic gap of graphene.  A compression along the surface normal direction of the QWs provides additional tunability for the topological gap, resulting in an enhanced gap up to 120 meV.  Our results lay a foundation for innovative designs of spintronic nanodevices.

\begin{acknowledgement}
Computation was carried out at HLRN Berlin/Hannover (Germany). L.K. acknowledges financial support by the Alexander von Humboldt Foundation of Germany. B.Y. and C.F. acknowledge financial support from the European Research Council Advanced Grant (ERC 291472). C.F.C. was partially supported by the DOE through the Cooperative Agreement DE-NA0001982.
\end{acknowledgement}

\begin{suppinfo}
The configurations of bilayer BiTeI-Graphene and the total energy variation as a function of the shifted distance are presented.  Also shown is an illustration of the effect of multilayers of BiTeI on the electronic properties of the QW and the associated mechanism.  The electronic band structures of the single- and multi-layer BiTeCl/Graph and BiTeBr/Graph QWs are also presented.
\end{suppinfo}


\end{document}